\documentclass[fleqn,12pt]{wlscirep}
\usepackage[utf8]{inputenc}
\usepackage[T1]{fontenc}
\usepackage{comment}
\usepackage{booktabs,array}
\usepackage{enumitem}
\usepackage{arydshln}
\usepackage{colortbl}
\usepackage{adjustbox}

\usepackage{subcaption}

\usepackage{appendix}

\title{Analysis of COVID-19 first wave in the US based on demographic, mobility, and environmental variables}
\author[1,*]{Dario Spiller}
\author[2]{Gabriele Santin}
\author[3,4]{Alessandro Sebastianelli}
\author[2,5,6]{Lorenzo Lucchini}
\author[2]{Riccardo Gallotti}
\author[7]{Brennan Lake}
\author[4]{Silvia Liberata Ullo}
\author[3]{Bertrand Le Saux}
\author[2]{Bruno Lepri}

\affil[1]{School of Aerospace Engineering, Sapienza University of Rome, Rome 00138, Italy}
\affil[2]{Bruno Kessler Foundation, Digital Society Center, Trento 38123, Italy}
\affil[3]{European Space Agency, $\Phi$-lab, Frascati 00044, Italy}
\affil[4]{University of Sannio, Benevento, 82100, Italy}
\affil[5]{Centre for Social Dynamics and Public Policy, Bocconi University, Milan 20100, Italy}
\affil[6]{Institute for Data Science and Analytics, Bocconi University, Milan 20100, Italy}
\affil[7]{Cuebiq Inc., New York, NY 10001, USA}

\affil[*]{dario.spiller@uniroma1.it}

\begin{abstract}
COVID-19 had a strong and disruptive impact on our society, and yet further analysis on most relevant factors explaining the spread of the pandemic are needed. Interdisciplinary studies linking epidemiological, mobility, environmental, and socio-demographic data analysis can help understanding how historical conditions, concurrent social policies and environmental factors impacted on the evolution of the pandemic crisis. This work deals with a regression analysis linking COVID-19 mortality to socio-demographic, mobility, and environmental data in the United States during the first half of 2020, i.e., during the COVID-19 pandemic first wave. This study can provide very useful insights about risk factors enhancing mortality rates before non-pharmaceutical interventions or vaccination campaigns took place. 
Previous analyses already pointed out the contribution of environmental variables and mobility, even though they tend to provide conflicting opinions. However, the relevance of socio-demographic inequalities has not been investigated in further detail, and it should be taken into account to get an adequate understanding of the pandemic evolution. 
Our cross-sectional ecological regression analysis demonstrates that, when considering the entire US area, the socio-demographic variables globally play the most important role with respect to environmental and mobility variables in describing COVID-19 mortality. Compared to the complete generalized linear model considering all socio-demographic, mobility, and environmental data, the regression based only on socio-demographic data provides a better approximation and proves to be a better explanatory model when compared to the mobility-based and environmental-based models. However, when looking at single entries within each of the three groups, we see that the mobility data can become relevant descriptive predictors at local scale, as in New Jersey where the time spent at work is one of the most relevant explanatory variables, while environmental data play contradictory roles. 
\end{abstract}
\begin{document}

% define new justified column with indentation
\newcolumntype{Y}[1]{%
  >{\small\raggedright\everypar{\hangindent=1.2em}\arraybackslash}p{#1}%
}

\flushbottom
\maketitle
\thispagestyle{empty}

\section{Introduction}
% generic intro on Covid19
Since late 2019, when it first appeared in Wuhan, China \cite{Li2020}, the new coronavirus disease (COVID-19) has drastically impacted our life and new variants continue to emerge and produce subsequent pandemic waves. As of the end of December 2022, it has caused more than 6 million official deaths in the world and more than 650 million confirmed cases \cite{WHO_web, Dong2020}. The scientific community started quite soon to investigate the new characteristics of COVID-19, starting from the transmission modalities up to finding the best ways to contrast it. As a consequence, a great number of papers have been published on this topic, investigating the problem from many perspectives and using different analysis approaches \cite{Adhikari2020,Zheng2020,Shereen2020,Chu2020,Yang2020,Rothan2020}. 

% correlation between Covid and mobility
Even though the vaccination campaign has been pursued globally, non-pharmaceutical interventions (NPIs) such as social distancing and face masks have continued to play a key role to mitigate the risk of new variants and subsequent pandemic waves \cite{Chu2020,DSleator2020}, as also discussed in recent studies \cite{Park_2022}. Some of the most effective approaches to mitigate the negative effects of the pandemic waves were strict social distancing, face mask use, minimization of social activities, contactless home delivery service, and self-quarantine rules \cite{Chen2020, Wellenius2021,Chernozhukov2021,Kwon2021,Wang2021}. As demonstrated by the early 2020 Wuhan (China) shutdown, which  has been associated with reductions in case incidence \cite{Tian2020}, it is worth analysing the effect of mobility and social interaction restrictions on COVID-19 transmission. 

Stay-at-home orders, workplace closures, and public information campaigns were effective
in decreasing the confirmed case growth rate \cite{Schlosser2020,Li2021}, even though the effect of social distancing on decreasing transmission is not likely to be perceptible for at least 9–12 days after implementation, and the delay might be longer \cite{Badr2020,Carteni2020,Cot2021}. In particular, the mitigation effects induced by the lockdown and the reduced mobility have been registered in several countries, e.g. in Italy \cite{Carteni2020}, in Japan \cite{Yabe2020}, in  the US \cite{Pan2020,lucchini2021living}, in China \cite{Kraemer2020}, and in worldwide analyses \cite{Oztig2020,Nouvellet2021}.  The role of other potential mitigating factors (e.g., wearing face masks and hand-washing, or seasonal changes \cite{Gatalo2021}) could also contribute to the decline in the case growth rate \cite{Badr2020}. 

A different line of studies has also reported a relationship with various environmental variables, suggesting that the spread of the COVID-19 disease is enhanced in colder and drier climates. However, evidence is still scarce and mostly limited to a few countries. For instance, a strong relationship is found when correlating the virus spread to mean temperature, minimum temperature, and atmospheric pressure in different Spanish provinces \cite{LocheFernandez-Ahuja2021}. %The most affected provinces were Soria, Segovia and Ciudad Real, which are the coldest. On the opposite side, places such as Southern Spain, the Baleares, and Canary Islands showed a lower rate of spread. This might be related to the warmer climate and the insularity of these islands. 
In Chile, there is evidence that COVID-19 transmission  was mostly related to three main climatic factors (minimum temperature, atmospheric pressure and relative humidity)\cite{Correa-Araneda2021}, with the greater transmission in colder and drier cities when atmospheric pressure was lower.
Also, results of analyses carried out using Korean data suggest that various environmental factors (e.g., duration of sunshine, ozone level, temperature) may play a role in COVID-19 transmission\cite{Lim2021}.
However, when moving to a global scale, the most shared feeling is that it remains unclear if the magnitude of the effect of temperature or humidity on COVID-19 is confounded by the public health measures implemented widely during the first pandemic wave\cite{Paraskevis2021}.  Indeed, demographic factors, together with other determinants, can also affect the transmission, and their influence may overcome the protective effect of climate, where favourable \cite{Spada2021}. With limited policy interventions, seasonal patterns of disease spread might emerge, with temperate regions of both hemispheres being most at risk of severe outbreaks during colder months. Nevertheless, containment measures play a much stronger role and overwhelm the impacts of environmental variation, highlighting the key role of policy interventions and non-pharmaceutical interventions (NPIs) in curbing COVID-19 diffusion within a given region. \cite{Ficetola2021,Baker2021,Mecenas2020,Poirier2020}.

Associations between COVID-19 cases and other confounding variables have been registered by several studies. For instance, social inequalities have an effect on the application of mobility policies, as disadvantaged groups cannot usually reduce their mobility as sharply as other more affluent social groups \cite{Gozzi2021, Chang2020,Zarro2022}. From a county-level study on COVID-19 cases and deaths in the US, comorbidities such as chronic lung diseases and cardiovascular diseases, demographic and social factors such as gender and age, environmental factors such as air pollution can explain the diffusion of the pandemic among particular social groups \cite{Vahabi2021}. Also, it has been shown that meteorological and air quality parameters were correlated to COVID-19 transmission in two large metropolitan areas in Northern Italy as Milan and Florence and in the autonomous province of Trento \cite{Lolli2020}.
The link between air pollutants and COVID-19 mortality has been the subject of many analyses in the recent literature. Some authors have considered the relationship between long-term exposure to air-borne particle matter (PM$_{2.5}$, PM$_{10}$) founding that higher historical exposures are positively associated with higher COVID-19 mortality rates \cite{Solimini2021,Wu2020}. Other studies focused on the correlation between COVID-19 cases and simultaneous or recent exposure to pollutants. As a general result, there are elements to confirm the existence of a link between pollution and the risk of death due to the disease, as it has been shown from studies focusing on Italy \cite{Accarino2021,Dettori2021,Zoran2020,Coccia2020}, the US \cite{Son2020,Razzaq2020}, England \cite{Travaglio2021}, and also on a world scale \cite{Ali2020,Singh2021,Sebastianelli2021b}.

Meteorological and human mobility data have been already put together in literature, for instance, they have been used to model the COVID-19 diffusion in Brazil \cite{DaSilva2021, Martins2020}, Asia \cite{Baniasad2021}, and Bangladesh \cite{Hassan2021}. Considering mobility and air pollution together for the Chinese area, a unit increase in the human mobility index was associated with a 6.45\% increase in daily COVID-19 confirmed cases, and the air quality index mediated 19.47\% of this association \cite{Zhu2020}.

% contributions
Dealing with the first wave of COVID-19 in the US, this paper focuses on the relationships among  mortality rates of COVID-19, environmental variables, mobility, and other socio-demographic confounding variables such as population distribution in terms of poverty or access to education. This research deals with a cross-sectional ecological regression analysis focusing at comparing the impact of the three groups of variables in describing COVID-19 mortality. A generalized linear model is chosen to find relationships between the number of deaths and the variable groups, both considering all socio-demographic, mobility, and environmental data, and also considering single groups or couples of data groups. This study reports both results at global US scale and results at local, state-level scale. In this way, we can underline how different variables can play different roles depending on the considered geographical scale. The presentation of the results mainly concerns the discussion of the $\beta$ coefficients of the generalized linear model, as they are direct indicators of the relevance and impact of predictor variables in explaining the number of COVID-19 deaths.

\section{Results}
\label{sec:results}
% NON FAREI RIFERIMENTO A PRIMA ONDATA, PERCHè LA PRIMA ONDATA PUò VARIARE DA PAESE A PAESE

% Area of interest and time window
The analyses described in this paper are focused on the contiguous United States (US) area, thus  excluding the states of Alaska and Hawaii and all the other offshore insular areas, such as American Samoa, Guam, the Northern Mariana Islands, Puerto Rico, and the US Virgin Islands. The geographical extent of the area under investigation allows us to consider many different environmental conditions, in this way accounting for the effects of climate and environmental variables. The temporal window of our analysis coincides with the first COVID-19 wave in the US. Specifically, our data cover the period going from December 31st, 2019, to August 8th, 2020. The temporal window is determined by the availability of mobility data (see Sec. \ref{sec:mob_data}), but allows us to perform an analysis which is not  affected by the vaccination campaign, thus maximizing the correlation between the COVID-19 mortality and the predictor variables taken into account.

% Le time series sono state convertite in dati puntuali???

% \begin{tabular}{p{40mm}p{40mm}p{40mm}p{40mm}}
% \toprule
% \multicolumn{4}{c}{Features}\\
% \multicolumn{2}{c}{Demography} & Mobility & Environment\\
% \midrule
%     \texttt{medianhousevalue} & \texttt{pct\_owner\_occ} &    \texttt{home\_time} &    \texttt{o3\_mean}\\
%     \texttt{medhouseholdincome} &  \texttt{pct\_blk} &  \texttt{work\_time} &    \texttt{o3\_std} \\
%     \texttt{education} & \texttt{hispanic} &   \texttt{other\_time} &    \texttt{relHum\_mean} \\
%     \texttt{beds} &   
%     \texttt{pct\_asian} & \texttt{travelled\_distance} &    \texttt{relHum\_std} \\
%     \texttt{obese} &  \texttt{pct\_native} &  \texttt{home\_time\_change} &    \texttt{specHum\_mean} \\
 %    \texttt{smoke} & \texttt{prime\_pecent} &   \texttt{work\_time\_change} &    \texttt{specHum\_std} \\
%     \texttt{mean\_pm25} &   \texttt{mid\_pecent} &  \texttt{other\_time\_change} &    \texttt{temp\_mean} \\
%     \texttt{poverty} & \texttt{older\_pecent} & &    \texttt{temp\_std} \\
% \bottomrule 
% \end{tabular}

\begin{table}[t!]
\resizebox{\textwidth}{!}{%
\centering
\begin{tabular}{Y{3mm}Y{40mm}Y{5mm}Y{38mm}Y{4mm}Y{28.5mm}Y{3mm}Y{33mm}}
\toprule
\multicolumn{8}{c}{\textbf{Features}}\\
\multicolumn{4}{c}{\emph{Socio-Demographic}} & \multicolumn{2}{c}{\emph{Mobility}} & \multicolumn{2}{c}{\emph{Environmental}}\\
\midrule
(D$_1$) &
Median house value & 
(D$_9$) & 
Owner-occupied housing  &
(M$_1$)&    
Home time  &
(E$_1$) &    
O$_3$ mean\\
(D$_2$) &
Median household income &
(D$_{10}$)&  
Black  &
(M$_2$)&  
Work time  &
(E$_2$)&    
O$_3$ std \\
(D$_3$) &
Less than high school  &
(D$_{11}$)& 
Hispanic  &
(M$_3$)&   
Other time  &
(E$_3$)&    
Rel. humidity mean \\
(D$_4$)  &
Rate of hospital beds  &
(D$_{12}$)&   
Asian  &
(M$_4$)& 
Travelled distance &
(E$_4$) &    
Rel. humidity std \\
(D$_5$)  &
Obese people  &
(D$_{13}$)&  
Native Americans &
(M$_5$) &  
Home time change & 
(E$_5$)&    
Spec. humidity mean \\
(D$_6$)  &
Smokers  &
(D$_{14}$)& 
15-44 years of age &
(M$_6$) &   
Work time change  &
(E$_6$)&    
Spec. humidity std \\
(D$_7$)  &
PM$_{2.5}$ long-term exposure  &
(D$_{15}$)&   
45-64 years of age  &
(M$_7$)&  
Other time change  &
(E$_7$)&    
Temperature mean \\
(D$_8$)  &
Below Poverty Line  &
(D$_{16}$)& 
$\ge$65 years of age  &
&  
&(E$_8$)&    
Temperature std \\
\bottomrule 
\end{tabular}}
\caption{Features used as predictors in our analyses, arranged by group.}
\label{tab:features}
\end{table}

%\paragraph{Preliminary analysis.}
Starting from the literature results summarized in the Introduction, we hypothesize that the mortality rates of COVID-19 in the US can be explained through socio-demographic, mobility, and environmental features. The groups of features are shown in Table~\ref{tab:features}, and a detailed description of the variables is reported in Sec. \ref{sec:data}. Socio-demographic data is  provided as cross-sectional data, i.e., only one single value is provided for each county over the entire time window. Hence, even though mobility and environmental data are provided as time series, we have averaged all the variables in the entire time window in order to be consistent with the socio-demographic information and perform a consistent cross-sectional regression. A brief description of the data used in this study is reported in Sec. \ref{sec:data_descr}, whereas a more detailed explanation can be found in Sec. \ref{sec:SI_data} of the Supplementary Material. Finally, a preliminary correlation analysis is reported in Sec. \ref{sec:corr_analysis} of the Supplementary Material, showing that there is no significant data correlation within each group of variables and they can be treated as independent predictors. 

This paper is intended to provide insights into the variables that can better explain the COVID-19 mortality rates in the US during the first wave using an ecological regression approach. Ecological regression is the statistical method of running regressions on aggregates, as in the case of this paper, where raster data was processed with zonal statistics on US counties and states. Potentialities and limitations of using  ecological regression methods on aggregate data are reported and well-known in literature\cite{Gelman2001,Robinson1950,Greenland1994}, and they were considered to critically discuss the relationship between  mortality rate and PM$_{2.5}$ long-time exposures\cite{Wu2020}.  Hence, the goal of this work is to help understand the most influential factors dictating the first pandemic wave in the US. It is also noteworthy that the proposed generalized linear model is not to be used to perform prediction over other time intervals, primarily because non-pharmaceutical interventions and vaccination campaigns strongly impacted the pandemic diffusion and the mortality rates and their effects are not included in our model.

% Linear models and FVU
% comparison between global and state-level models
\subsection{Explanatory role of features' groups} 

\begin{figure}[t!]
    \centering
    \includegraphics[width=.7\textwidth]{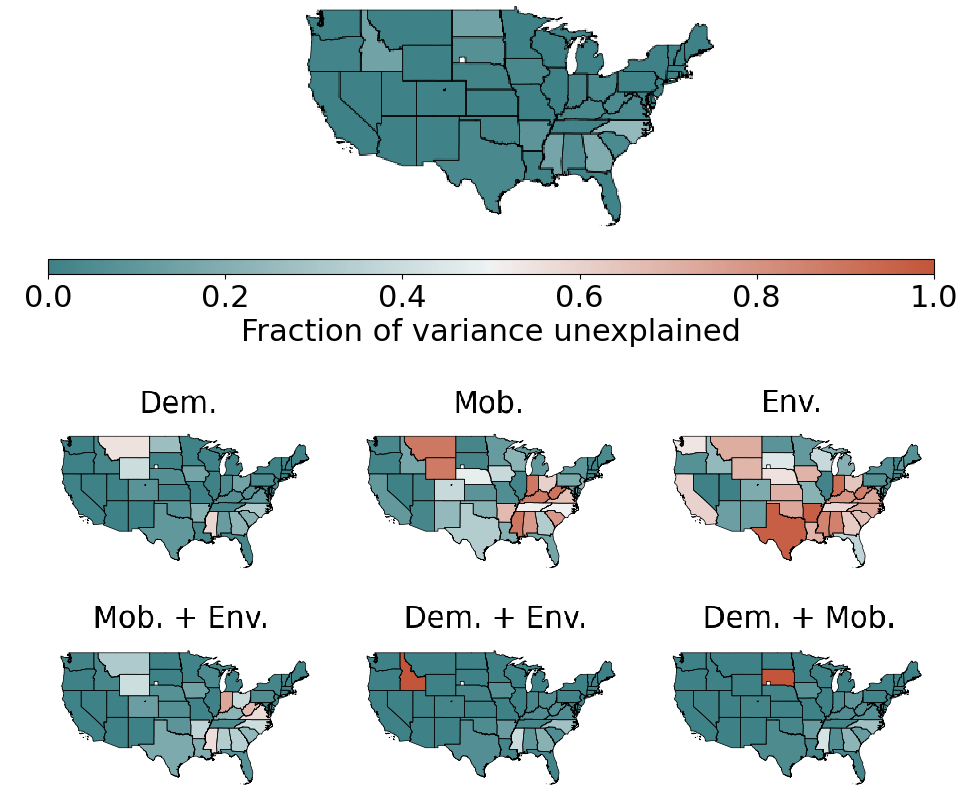}
\caption{Fraction of Variance Unexplained (counties are grouped by state). The figure reports the FVU index considering all the three gropus of features (top), a single group of features (middle), and group pairs of features (bottom). A strong explanation is described by FVU values close to zero (green color), whereas weak results correspond to values close to one (red color).}
    \label{fig:maps_fvu_states}
\end{figure}

As in other similar studies \cite{Wu2020}, we fit a negative binomial mixed model (NBMM) where the outcome is identified as the logarithm of COVID-19 mortality rates in US during the first pandemic wave. The predictor variables are the ones included in the socio-demographic, mobility, and environmental data groups. Differently from other studies where a unique global model was developed \cite{Wu2020}, in this work we have obtained a specific model for each state in the contiguous United States area using the data corresponding to the internal counties for the fitting.  Additional information about NBMM  is reported in Sec. \ref{sec:data} and in the Supplementary Material, along with the comparison of the results related to different fitting strategies. 
Fig. \ref{fig:maps_fvu_states} shows the results of the NBMM fitting in terms of Fraction of Variance Unexplained (FVU) (the FVU values shown in the figure are reported in Table \ref{tab:FVU} in Sec. \ref{sec:table_map} of the Supplementary Material). In the context of a regression task, FVU is the fraction of variance of the dependent variable (COVID-19 mortality rates) which cannot be explained, i.e., which is not correctly predicted by the  explanatory variables. Hence, the best regression results are provided when FVU is close to zero.

The obtained results demonstrate how relevant the socio-demographic variables are as compared to the other two variables' groups. Indeed, looking at the results from the single models, the FVU values of the socio-demographic model are always lower than the ones from the mobility and environmental models. Moreover, compared with the FVU results related to the complete model (i.e., the plot on the top), the difference in explanation capability is far lower for the socio-demographic group with respect to mobility and environmental ones. Indeed, the socio-demographic map is almost completely green (apart from Montana and Mississippi), whereas the mobility and the environmental maps report higher values and show many red states (FVU close to 1).

To better understand the relative explanatory relevance, three mixed models, each one based on group pairs of features, are analysed as well. More specifically, the models based (i) on socio-demographic and mobility data (\emph{Dem+Mob}), (ii) on socio-demographic and environmental data (\emph{Dem+Env}), and (iii) on mobility and environmental data (\emph{Mob+Env}) are defined and shown in the last three maps in the bottom of Fig. \ref{fig:maps_fvu_states} (they are also reported in their relative columns of Table \ref{tab:FVU} of the Supplementary Material). Also, in this case, it can be well appreciated that the models including the socio-demographic variables perform better than the model having only mobility and environmental data. 

\subsection{Analysis of the coefficients in the global generalized linear model}
\label{sec:beta_global_model}

By comparing how features play different roles in different states, we can understand how general results considering the entire US area can be mapped into individual states. Hence, we turn our attention to the standardized $\beta$ coefficients that reveal how features correlate with COVID-19 mortality in each US state. The following analysis is performed considering the complete model with all the variables in the three groups (the column \emph{All} in Table \ref{tab:FVU} of the Supplementary Material). To properly compare the coefficients and consider the variability of the number of deaths across the states, all the coefficients for every single state have been normalized such that they sum to one. In this way, we can consistently compare $\beta$ coefficients in states with a high number of deaths (which have quite high values before the normalization) with $\beta$ coefficients in states with a small number of deaths (which have small values before the normalization).  

\begin{figure}[!b]
    \centering
    \hspace*{-1.0cm}
    \includegraphics[width=1.1\textwidth]{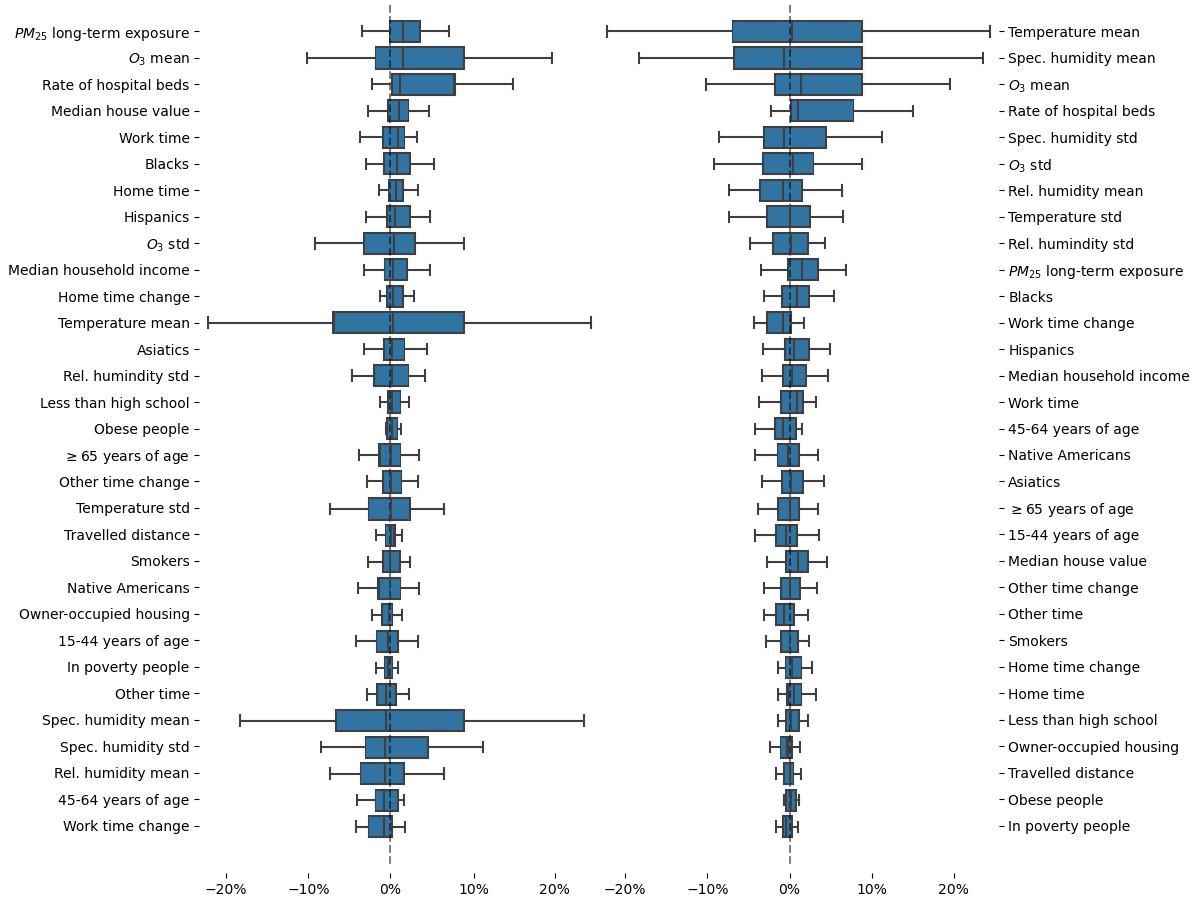}
\caption{Generalized Linear Model’s $\beta$ coefficients showing the different impact of socio-demographic, mobility, and environmental features. Variables are sorted with decreasing median on the left and decreasing interquartile ranges on the right.}
    \label{fig:beta_coeffs_new}
\end{figure}

The results of this analysis are depicted in Fig. \ref{fig:beta_coeffs_new} using box plots displaying interquartile ranges (the boxes reporting the spread difference between the 75th and 25th percentiles of the data). In Fig. \ref{fig:beta_coeffs_new}, the variables have been sorted with decreasing median on the left and decreasing interquartile ranges on the right. It is noteworthy that the $\beta$ coefficients greatly vary across the states, as one can appreciate by observing that almost all the reported variables present both positive and negative correlations with COVID-19 mortality. 

The output of this analysis demonstrates that it is not straightforward to find  commonalities in all the states, i.e. there are no variables which  correlate always positively or negatively with COVID-19 mortality in all the US states. There are, in fact, just a few variables which tend to correlate only positively or negatively, as the PM$_{2.5}$ and the rate of hospital beds, where the box only covers positive values, and the work time change (i.e., change of time spent at work with respect to the baseline value computed in the month frame from January 3rd, 2020, to February 6th, 2020 - further details are reported in the Supplementary Material), where the box only covers negative values. However, there are no strong positive or negative dependencies, and the conclusions of this analysis may be summarized in the following paragraphs.

\paragraph{Variables with similar impacts on mortality rates.} Many variables have a small box plot centred in zero. Considering the normalization of the $\beta$ coefficients, this outcome means that, on average, almost all the variables tend to have comparable impacts on the number of deaths within each state.

\paragraph{Historical exposure to PM$_{2.5}$.} Long exposures to PM$_{2.5}$ seem to increase the probability to die with COVID-19. Indeed, from the right box plot in Fig. \ref{fig:beta_coeffs_new}, the  PM$_{2.5}$ long-term exposure variable has the higher median value among all the variables, and the interquartile range contains all positive $\beta$ coefficients.
This result is consistent with previous studies \cite{Wu2020}, and it gives credence to the opinion that previous medical and health conditions can heavily influence the response to COVID-19 regardless to other concomitant factors.   
This result may help policymakers to take
precautionary measures for the areas with
historical high values of PM$_{2.5}$ and provides a solid scientific argument for lowering the U.S. National Ambient Air Quality Standards for PM$_{2.5}$.

\paragraph{Effectiveness of stay-at-home restrictions.} Looking at the mobility variables, it is not easy to draw clear conclusions on the global effectiveness of the stay-at-home restrictions. Indeed, analysing the outcome in Fig. \ref{fig:beta_coeffs_new}, only some mobility variables provide results consistent with the expectations, such as the \emph{work time} (i.e., the time spent at the workplace), which has a positive median value of the $\beta$ coefficient. Hence, this result suggests that spending more time at work, so interacting with people at work, could increase the number of deaths due to COVID-19. However, this analysis on the entire US area also says that \emph{home time} (i.e., the time spent at home, hence only interacting with  people from your own family) is positively correlated with the number of deaths, so that staying at home does not seem to be an effective alternative to staying at the workplace. Similarly, a difficult interpretation of the result is related to the \emph{work time change} variable, which is defined as the difference between the work time during the COVID-19 pandemic and the work time in pre-COVID conditions (which is a negative variable, as one can understand from Figure~\ref{fig:time_series_work_time} in Sec.~\ref{sec:time_evolution_wt} - further details are provided in Sec. \ref{sec:mob_data}). Indeed, \emph{work time change} is negatively correlated with COVID-19 deaths, which slightly contradicts the result about work time. This result suggests that work time change (as well as home time change) might be better interpreted as a predictand instead of a predictor. Following this interpretation, work time change should be seen as a consequence of the high number of deaths, and we are not able to read the effects of this social policy from our numerical results. As a result, it is not easy to provide an overall interpretation of the mobility variables impact. The outcome of this analysis is that mobility variables do play a relevant role, but it is difficult to provide a uniform and consistent interpretation of the results when considering the entire US area. This point is addressed also in Sec. \ref{sec:state_examples} when discussing the impact of mobility in the New Jersey state. 

\paragraph{Low explanation ability of environmental data.} Considering the global model, it turns out that the maximum values of the $\beta$ coefficients are attained by environmental variables (mean temperature, mean specific humidity, and mean O$_3$). This is not in contrast with the previous analysis and should not be read as a contradictory outcome. Indeed, as explained in the following Section \ref{sec:state_examples}, these three environmental variables have quite a spread and disordered distribution, which explains the high interquartile range and does not provide any meaningful relationship between these variables and COVID-19 deaths. Indeed, considering the model based only on environmental data, its explanatory ability is drastically reduced if compared to the complete model.

\subsection{Detailed analysis over a selection of significant states}
\label{sec:state_examples}

This state-level analysis provides some interesting elements of discussion with respect to the FVU analysis reported in the previous section, strengthening the interpretation of the results with some specific examples.

\begin{figure}[!b]
    \centering
    \includegraphics[height=0.7\textheight]{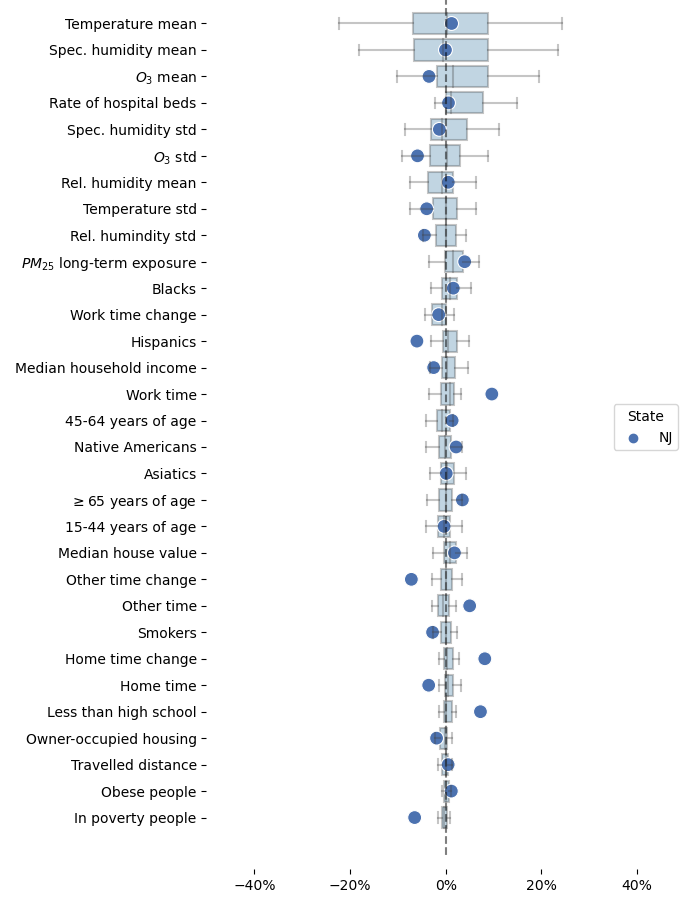}
\caption{Generalized Linear Model’s  reporting New Jersey $\beta$ coefficients with blue dots over the global US trend.}
    \label{fig:beta_coeffs_NJ}
\end{figure}

\paragraph{Role of mobility variables in New Jersey.} If on one hand, we can state that the socio-demographic variables can explain quite well the COVID-19 mortality across the entire US, on the other hand, this is not true when considering a state-level analysis. Mobility variables can provide further information at the state level (even though there is not a uniform impact on the US area, as discussed in Sec. \ref{sec:beta_global_model}), as in the case of \emph{home time}, \emph{work time}, and \emph{other time}, which are the time spent at home, the time spent at work, and the time spent in other activities (not including travelling). From Fig. \ref{fig:beta_coeffs_NJ} we can notice that the mobility variables \emph{home time}, \emph{other time}, and \emph{work time} attain very high  $\beta$ coefficient in New Jersey (NJ). We can attempt to explain this result by considering that, during the first wave, New York City (NYC) was one of the most hit cities, and a lot of people were commuting from NYC to New Jersey, first for working reasons and then to escape from the dangerous New York outbreak. Specifically, it is worth noting that work time is positively correlated with COVID-19 deaths, whereas home time is negatively correlated. This means that, for the specific case of New Jersey, a lower number of deaths was expected spending more time at home and less time at the workplace. 

\begin{figure}[!th]
     \centering
     \includegraphics[width=1\columnwidth]{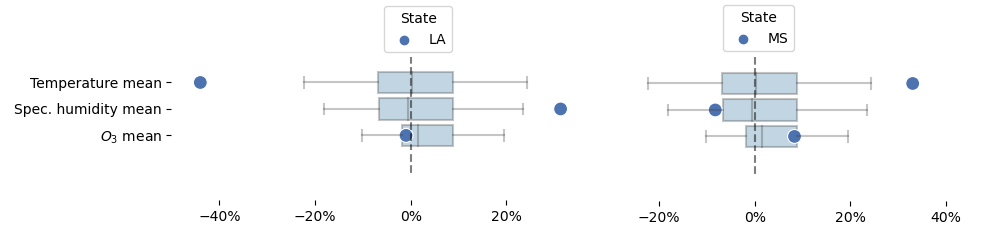}
     \caption{Generalized Linear Model’s  reporting Louisiana versus Mississippi $\beta$ coefficients for mean values of temperature, specific humidity, and O$_3$.}
     \label{fig:LA_vs_MS}
     \includegraphics[width=1\columnwidth]{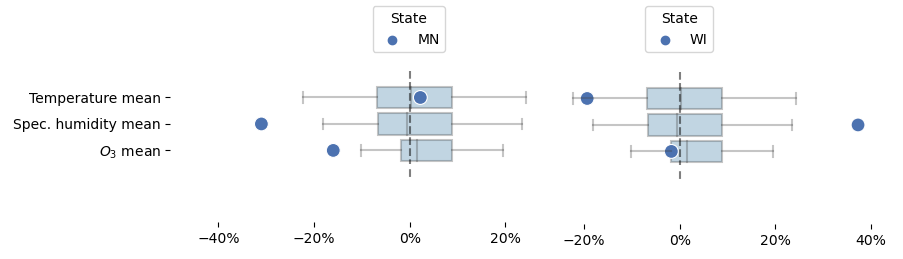}
     \caption{Generalized Linear Model’s  reporting Minnesota versus Wisconsin $\beta$ coefficients for mean values of temperature, specific humidity, and O$_3$.}
     \label{fig:MN_vs_WI}
\end{figure}

\paragraph{Contradictory environmental influence in southern states.} As already discussed, environmental variables do not explain COVID-19 deaths as well as the other two families of variables. This result can be appreciated in Fig. \ref{fig:LA_vs_MS} where the Louisiana and Mississippi mean environmental variables are compared. Being two neighbouring states, one would expect the environmental variables to behave in a similar and consistent manner. However, what we register is that both the mean temperature value and the mean specific humidity have very contrasting $\beta$ coefficient values, as the temperature attains a large negative value in Louisiana and a large positive value in Mississippi. Similarly, the specific humidity positively influences the number of deaths in Louisiana, whereas a positive impact is registered in Mississippi. 

\paragraph{Contradictory environmental influence in northern states.} As in the previous point, a contradictory influence of the environmental variables is noted also in the northern states. In Fig. \ref{fig:MN_vs_WI} we compare Minnesota and Wisconsin $\beta$ coefficient values. Also this time we can notice opposite influences of the variables in the two neighbouring states, particularly evident for the specific humidity, where we can see a strongly negative outlier in Minnesota and a strongly positive outlier in Wisconsin. 

\section{Discussion}

In this paper, we modelled the COVID-19 mortality rate as a function of socio-demographic, mobility, and environmental variables. We have used a cross-sectional ecological analysis based on a generalized linear model. The goal of this paper was to analyse how the three groups of variables correlated with COVID-19 mortality rates. Hence, different models have been fitted for each US state using data for each internal county as predictors. A complete model based on the complete set of variables has been compared to models fitted on single groups (i.e., socio-demographic model, mobility model, and environmental model) and on mixed combinations of groups (i.e., socio-demographic and mobility model, socio-demographic and environmental model, mobility and environmental model).

 % Discussion about the great relevance of demographic data
 When considering the fraction of unexplained variance and the three separated groups of variables, the best approximation of the complete model is provided by using only socio-demographic data. Hence, from a global point of view, the mortality impact of the first wave in US was strongly guided by population distribution in terms of social and wealth inequalities, ethnic components, and age composition.
% Discussion on Mobility
 The second most relevant variable is mobility, especially for some states (such as New Jersey) where during the first wave there has been a lot of moving people which influenced the evolution of the epidemics and the related mortality rates. 
 % Results of mixed models
 When considering the mixed combination of variables groups, then models containing socio-demographic data perform better than the model based only on mobility and environmental data. 
 % aggiunto per i commenti di Gallotti
 It is worthy to comment that these results could depend on the temporal period chosen for this study. Indeed, this analysis shows that, during the first pandemic wave, socio-demographic data (i.e. historical conditions) better explain the mortality rates with respect to the other predictors. However, environmental and mobility data could play different and more relevant roles analysing the spread of COVID-19 during a longer time frame (which is beyond the purpose of this study). This is particularly true when focusing on some mobility variables, such as home time change or work time change, which seem to better reflect policy interventions taken as consequence of the pandemic spread rather than predictors (and causes) of the mortality rates. 

% Comment on PM2.5 and environmental
The strong influence of long-time exposures to PM$_{2.5}$ already noticed in \cite{Wu2020} has been confirmed, whereas no relevant impact of the environmental variables has been noticed. On the contrary, two examples have been reported for neighbouring states in the North and in the South of the US where environmental variables play completely opposite roles, meaning that similar environmental conditions do not correlate with a similar number of COVID-19 deaths as other variables have a stronger influence. 
This result is in line with other previous studies where no clear relationships between environmental factors and COVID-19 mortality rate were found. Specifically, this study confirms that the effect of weather and climate variables cannot be excluded, however, under the conditions of the first pandemic wave, it might be difficult to be uncovered, as already pointed out by Paraskevis et al. \cite{Paraskevis2021}. Even though seasonal patterns of disease spread might emerge using other analysis methodologies or input data, containment measures play a much stronger role and overwhelm the impacts of environmental variation, highlighting the key role of policy interventions and NPIs in curbing COVID-19 diffusion within a given region \cite{Ficetola2021,Mecenas2020}. However, it is worth reminding that there are other works claiming a strong relationship between environmental variables and COVID-19 mortality rate \cite{LocheFernandez-Ahuja2021, Correa-Araneda2021,Lim2021,ISLAM2022e10333}. %Recent research results highlight that COVID-19 could have a seasonality similar to the seasonality of influenza-like illnesses \cite{COCCIA2022112711}, even though a widespread and shared opinion is that environmental impact on COVID-19 still shows spatial heterogeneity and uncertainty, different environmental factors interactively affect COVID-19 incidence, and the interactions with other factors better explain the pandemic dynamic transmission \cite{HAN2023933}.

 % Limitations of ecological analysis 
 As already denoted in other similar works \cite{Wu2020}, it is noteworthy that confusion between ecological associations and individual associations may present an ecological fallacy. In extreme cases, this fallacy can lead to associations detected in ecological regression that do not exist or are in the opposite direction of true associations at the individual level. However, ecological regression analyses still allowed us to make conclusions at the state level, which was useful to extract general tendencies. 

 % Limitations of cross-sectional regression
This paper investigated the first COVID-19 wave in the US using a cross-sectional regression, which does not take into account the time distribution of influencing variables and mortality rates. Even though this should not impact the high-level correlations, we may miss some details related to the time distribution of relevant confounders. In further detail, we cannot assess if, during the considered time window, some variables increased or decreased in predicting ability. This is particularly important for those variables which can change at relatively high frequency, i.e., mobility and environmental data. Hence, we can only capture an overall result to compare variable groups, but detailed results correlating time variations of mobility and environmental data with COVID-19 mortality rates is beyond the scope of the proposed model. 
Other considerations that are worthy to be mentioned relate to the precision of the mortality rates data, as there are no certified guarantees that all mortality data were collected in the same way and with the same procedures throughout all the US states. In the same fashion, mobility data could not describe unequivocally different conditions related to social interactions. For instance, working environments may have very different numbers of workers or different proximity conditions among workers, as well as the interpretation of the time spent at home is different for people living alone or people living in numerous families.

To conclude, this paper shows the complexity of modelling the COVID-19 mortality rate as a function of a small number of predominant confounding variables. As demonstrated with the previous example cases, there are some areas where mobility played a relevant role, while the environmental variables seem to have contradictory impacts. The long-term exposure to PM$_{2.5}$ remains a relevant factor as already pointed out in previous analysis \cite{Wu2020}. Considering groups of variables in the socio-demographic, mobility, and environmental data, the first subset is the most relevant and can better explain the COVID-19 mortality rate during the first wave in the US.

\section{Data and Methods}
\label{sec:data}

\subsection{Region of interest}
The basic geographical area used in this work is the county. Official boundaries of the US counties on a 1 : 500000 scale have been obtained from the United States Census Bureau \cite{Census_US}.
%\footnote{\href{https://www.census.gov/geographies/mapping-files/time-series/geo/cartographic-boundary.html}{https://www.census.gov/geographies/mapping-files/time-series/geo/cartographic-boundary.html}}. 

These data describe the boundaries of $3233$ counties, including city counties. 
% Out of these data, only $3142$ counties are kept for the analysis, and $91$ counties have been removed. These include the $78$ counties in Puerto Rico, the three counties in the Virgin Islands, the five counties in the American Samoa, the county of Guam, and the four counties in the Northern Mariana Islands.
Out of these data, only $3220$ counties are kept for the analysis, and $13$ counties have been removed. These include the three counties in the Virgin Islands, the five counties in the American Samoa, the county of Guam, and the four counties in the Northern Mariana Islands.

\subsection{Datasets}
\label{sec:data_descr}
A concise description of the data used in this work is reported, providing details for each group of variables. More information are provided in the Supplementary Material.

\subsubsection{Epidemiological data}
\label{sec:dengue_data}
The time series recording the number of deaths have been obtained from the COVID-19 Data Repository by the Center for Systems Science and Engineering (CSSE) at Johns Hopkins University\cite{Dong2020}. Also in this case, the extraterritorial counties that have been discarded from the counties' list are not considered. 
Additionally, we did not consider the records corresponding to non-geographical areas, i.e., two ships (Grand Princess and Diamond Princess) and two prisons (Federal Correctional Institution (FCI) and Michigan Department of Corrections (MDOC)).

Moreover, the Johns Hopkins data set contains some additional data that requires further pre-processing as follows:
\begin{itemize}
    \item For each state the tables contain entries that are marked as "Unassigned" and "Out of" (e.g., "Unassigned Utah" and "Out of Utah"). Even though ignoring these data may lead to an underestimation of the spread of the epidemic\footnote{\href{https://www.linkedin.com/pulse/county-less-covid-19-how-unassigned-out-state-counties-troy-hughes/}{https://www.linkedin.com/pulse/county-less-covid-19-how-unassigned-out-state-counties-troy-hughes/}}, we do not consider them since they are not geo-located at the county level and thus are out of scope for our analysis.

    \item There are some counties that appear in the data set at an aggregated level. In this case, the data reported for the aggregated territory is equally split between the single counties composing the aggregation. These are the cases in details:
    \begin{itemize}
        \item Dukes, Massachusetts, and Nantucket, Massachusetts, for which a unique entry of "Dukes and Nantucket, Massachusetts" is present.
        \item Kansas City, which is not a county, and whose territory overlaps with the four counties of Jackson, Clay, Cass, Platte (Missouri). 
        \item The state of Utah, for which the cases are reported at the jurisdiction level. There are $13$ jurisdictions\footnote{\href{https://coronavirus-dashboard.utah.gov/}{https://coronavirus-dashboard.utah.gov/}}, corresponding both to a single county or to multiple ones.
    \end{itemize} 
\end{itemize}
After these pre-processing steps, we have a time series of deaths for each of the $3220$ counties, for the time frame from January 22th, 2020 to August 8th, 2020.

\subsubsection{Socio-demographic data}
\label{sec:social_data}

\begin{table}[!t]
\centering
%\begin{tabular}{llp{3cm}p{5cm}}
\begin{tabular*}{\textwidth}{l @{\extracolsep{\fill}} lp{3cm}p{5cm}}
\toprule
ID &Name & Unit & Description\\
\midrule
D$_1$&Median house value&$1,000$& County-level median house value\\
D$_2$&Median household income&$1,000$& County-level median household income\\
D$_3$&Less than high school&dimension-less& Percent of the adult population with less than high school education\\
D$_4$&Rate of hospital beds&per population units of 100,000 people& Number of hospital beds per unit population\\
D$_5$&Obese people&dimension-less& Percent of the population with obesity\\
D$_6$&Smokers&dimension-less& Percent of current smokers\\
D$_7$&PM$_{2.5}$ long-term exposure&$\mu$g/m$^3$& County-level long-term exposure to PM$_{2.5}$ (averaged from 2000 to 2016)\\
D$_8$&Below Poverty Line&dimension-less& Percent of the population living below the poverty line\\
D$_9$&Owner-occupied housing&dimension-less& Percent of owner-occupied housing\\
D$_{10}$&Black&dimension-less& Percent of Black residents\\
D$_{11}$&Hispanic&dimension-less& Percent of Hispanic residents\\
D$_{12}$&Asian&dimension-less& Percent of Asiatic residents\\
D$_{13}$&Native americans&dimension-less& Percent of Native Americans residents\\
D$_{14}$&15-44 years of age&dimension-less& Percent of the population with 15-44 years of age\\
D$_{15}$&45-64 years of age&dimension-less& Percent of the population with 45-64 years of age\\
D$_{16}$&$\ge$65 years of age&dimension-less&Percent of the population with $\ge$65 years of age\\
\bottomrule
\end{tabular*}
\caption{Summary of the demographic variables, retrieved from Wu et al.\cite{Wu2020}.}
\label{tab:data_demography}
\end{table}

Socio-demographic data was retrieved from the work of Wu et al.\cite{Wu2020}, where a complete description of the dataset is reported. However, we used a slightly reduced number of variables (16 instead of 18), which are the ones already reported in Table \ref{tab:features} and listed with more details in Table \ref{tab:data_demography}.
The two variables have been removed to simplify the model, as they are redundant: the variable representing the percent of white population adds to 100\% with the other ethnicity percentages, and the variable representing the percentage of the population under 15 years old adds to 100\% with the other percentages of age range.
These variables were collected from the 2000 and 2010 Census, the 2005–2016 American Community Surveys and the 2009–2016 US Centers for Disease Control and Prevention (CDC) Compressed Mortality File. In further detail, the proportion of residents with obesity and the proportion of residents that are current smokers were evaluated from the Robert Wood Johnson Foundation’s 2020 County Health Rankings.
The county-level information on the number of hospital beds available in 2019 was retrieved from Homeland Infrastructure Foundation-Level Data (HIFLD). %It is worth noting that we selected only a portion of the data used in Wu et al.\cite{Wu2020}, as environmental data was collected here in an alternative way using Copernicus data, and mobility was not considered.

\subsubsection{Mobility data}
\label{sec:mob_data}
The mobility data consists of GPS-located trajectories provided for research purposes by Cuebiq Inc. within their ``Data for good'' program (\url{https://www.cuebiq.com/about/data-for-good/}).
Cuebiq collects data from devices that have opted-in to anonymized data sharing through a GDPR and CCPA-compliant framework. In addition to anonymizing data, the data provider obfuscates home locations at the Census Block Group level and removes sensitive points of interest from the dataset in order to preserve privacy.
Individual data are provided over the entire US territory with an approximate coverage of about 1\% of the national population. Individual trajectories are associated with a specific county based on the most visited location during nighttime and/or during weekends over the entire span of the data. Trajectories are processed using state-of-the-art techniques to obtain individual stop locations (for more details see Lucchini et al.~\cite{lucchini2021living}). For each county, only stop locations that fall into a county's area are included when computing county-specific mobility indicators. County areas are constructed by geo-hashing the stops on a regular two-dimensional grid with $600$m spatial steps in both dimensions (latitude and longitude). Whenever a rectangle overlaps with a boundary, all the stops falling within its borders are included in the neighbouring counties. This approach slightly overestimates the number of records, but it makes the intersection more robust.

The data are aggregated by day, so that we have one data point per county, per day, in the time frame from December 31st, 2019, to August 8th, 2020.

\begin{table}[!t]
\centering
%\begin{tabular}{llll}
\begin{tabular*}{\textwidth}{l @{\extracolsep{\fill}} lll}
\toprule
ID & Name & Unit & Description\\
\midrule
M$_1$ & Home time &hours&Time spent at home\\
M$_2$ & Work time &hours&Time spent at work\\
M$_3$ & Other time &hours&Time spent not at home and not at work\\
M$_4$ & Travelled distance &meters&Travelled distance\\
M$_5$ & Home time change &hours& Change of home time w.r.t. the baseline\\
M$_6$ & Work time change &hours& Change of work time  w.r.t. the baseline\\
M$_7$ & Other time change &hours& Change of other time w.r.t. the baseline\\
\bottomrule
\end{tabular*}
\caption{Summary of the mobility variables. The baseline values are computed as mean values in the month frame from January 3rd, 2020, to February 6th, 2020.}
\label{tab:data_mobility_reduced}
\end{table}

All the mobility indicators included in our analysis are computed as county-wide averages. First, individual patterns are aggregated by county of residence, second, derived metrics are computed. In particular, changes in mobility indicators are computed with respect to a baseline period ranging from 2020-01-03 to 2020-02-06 (extremes included). Within the baseline, period median mobility values are extracted for each different ``day of the week'' to account for periodic weekly patterns. 
Thus, at a specific date $t$, corresponding to a day of the week $d$, the change of the mobility indicator, $C_m(t|d)$ is computed as the relative difference between the median value of the indicator during the baseline period overall $d$-days of the week: $$C_m(t|d) = \frac{m(t|d) - Median[m(t'|d)]}{Median[m(t'|d)]},$$
where 2020-01-03$\le t' \le$ 2020-02-06.
The baseline period is selected to avoid all policy activations across the entire US territories (in this period, no policy had yet been introduced in the US for the control of the pandemic). 
Table \ref{tab:data_mobility_reduced} reports the mobility variables used for the analysis.

\subsubsection{Environmental data}
\label{sec:era5_data}

\begin{table}[!t]
\centering
\begin{tabular*}{\textwidth}{l @{\extracolsep{\fill}} lll}\toprule
ID & Name & Unit & Description\\
\midrule
E$_1$ & O$_3$ mean & dimension-less& Ozone mass mixing ratio at 1000 hPa\\
E$_2$ & O$_3$ std & dimension-less& Standard deviation related to the mean value of E$_1$\\
E$_3$ & Relative humidity mean &dimension-less&Relative humidity at 1000 hPa\\
E$_4$ & Relative humidity std &dimension-less&Standard deviation related to the mean value of E$_3$\\
E$_5$ & Specific humidity mean &dimension-less&Specific humidity at 1000 hPa\\
E$_6$ & Specific humidity std &dimension-less&Standard deviation related to the mean value of E$_5$\\
E$_7$ &Temperature mean &Kelvin&Air temperature at 1000 hPa\\
E$_8$ & Temperature std &dimension-less&Standard deviation related to the mean value of E$_7$\\
\bottomrule
\end{tabular*}
\caption{Summary of the environmental variables retrieved from the Climate Data Store of the Copernicus programme \cite{CDS_copernicus}.}
\label{tab:atmospheric_data}
\end{table}

Environmental data used for this study include temperature, humidity (relative and specific), and ozone mass mixing ratio, and they are summarized in Table \ref{tab:atmospheric_data}. Data have been retrieved by the ERA5 reanalysis dataset which can be freely accessed online from the Climate Data Store \cite{CDS_copernicus} and are elaborated by the European Centre for Medium-Range Weather Forecasts (ECMWF) as part of the Copernicus programme. 

Depending on the variables of interest, ERA5 data are provided at regular 0.25 degree $\times$ 0.25 degree grids or 0.1 degree $\times$ 0.1 degree grids. Hence, to assign a single value per day for each US county, a pre-processing phase is needed in order to 1) elaborate a finer mesh at a regular 0.01 deg $\times$ 0.01 deg grid using bilinear interpolation, and 2) group data within each county and average the values. Indeed, as for the mobility data, information are aggregated once per day in order to have a data point per county, per day, in the time frame from December 31st, 2019, to August 8th, 2020.
All the variables are measured at 1000 hPa. For each variable in the environmental dataset, mean value and standard deviation have been considered as explanatory variables in the regression analysis.

\subsection{Generalized linear model}
\label{sec:glm}
The results of this work have been obtained fitting a generalized linear model (GLM), specifically a negative binomial mixed model (NBMM) using COVID-19 deaths  as the outcome and all the data described in Sec. \ref{sec:data_descr} as predictors.

The NBMM approach has been implemented in language R by using the lme4 package \cite{JSSv067i01}, where all potential predictive variables are centered
and scaled prior to the analysis. 
Letting $\mathbb{E}$ denote an expected value, $p_i^{(dem)},\;p_j^{(mob)},\;p_k^{(env)}$ be the generic predictors in the socio-demographic, mobility, and environmental groups, respectively, weighted with associated coefficients $\beta_i^{(dem)},\;\beta_j^{(mob)},\;\beta_k^{(env)}$, the main model takes the following form: 
\begin{equation*}
    \log{\mathbb{E}(\text{COVID-19 \;deaths})} = \beta_0 + \sum_{i=1}^{N_{dem}}\beta_i^{(dem)} p_i^{(dem)}
    + \sum_{j=1}^{N_{mob}}\beta_j^{(mob)} p_j^{(mob)}
    + \sum_{k=1}^{N_{env}}\beta_k^{(env)} p_k^{(env)}
\end{equation*}
Magnitude and signs of the $\beta$ coefficients have been used in this research to extract information about the relationships between predicting variables and number of deaths.
%%%%%%%%%%%%%%%%%%%%%%%%%%%%%%%%%%%%%%%%%%%%%%%%%%%%%%%%%%%%%%%%%%%
\section*{Acknowledgements}
L. L. has been supported by the ERC project ``IMMUNE'' (Grant agreement ID: 101003183). We acknowledge the support of Simone Centellegher and Marco De Nadai in writing, optimizing, and scaling the code for the mobility data processing.

\section*{Author contributions statement}
D. S., G. S., B. L. S. and B. Lepri designed research and experiments;
D. S. analysed environmental data, performed research, summarized the findings and wrote the manuscript; 
A. S. analysed environmental data and performed experimental analyses;
G. S. performed experimental analyses and contributed to the writing of the manuscript;
L. L. preprocessed the mobility data;
B. Lake provided mobility data;
S. U., B. L. S., and B. Lepri supervised the different phases of the project.
D. S., G. S., A. S., L. L., R. G., B. L., S. U., B. L. S., and B. Lepri provided feedback, read, reviewed and approved the final manuscript.

\section*{Additional information}
\subsection*{Competing interests}
The authors declare no competing interests.

\subsection*{Data and source code Availability}
Mobility data from Cuebiq can be accessed only through the Data for Good initiative of the company\footnote{\url{https://www.cuebiq.com/about/data-for-good/}}. Limitations apply to the availability of this data, due to the rigorous anonymity constraints. All other data sources are freely available on the Internet. The number of COVID-19 deaths was obtained from the  Data Repository by the Center for Systems Science and Engineering (CSSE) at Johns Hopkins University\cite{Dong2020}. Socio-demographic data was retrieved from the work of Wu et al.\cite{Wu2020}, and environmental data was retrieved from the ERA5 reanalysis dataset which can be freely accessed online from the Climate Data Store \cite{CDS_copernicus}. Further details on each source's availability are reported in Section~\ref{sec:data_descr}.
% Code can be shared, upon reasonable request, contacting the corresponding author.
The code to replicate all the
experiments will be released in a public repository upon the first
review step.

\bibliography{biblio}
%%%%%%%%%%%%%%%%%%%%%%%%%%%%%%%%%%%%%%%%%%%%%%%%%%%%%%%%%%%%%%%%%%%

\label{MainLastPage}
\newpage
\clearpage 
\rfoot{\small\sffamily\bfseries\thepage/\pageref{LastPage}}%
\pagenumbering{arabic}
\setcounter{page}{1}
\appendix
\appendixpage

\renewcommand{\thesection}{S\arabic{section}}  
\renewcommand{\thetable}{S\arabic{table}}  
\renewcommand{\thefigure}{S\arabic{figure}}

\setcounter{figure}{0}
\setcounter{table}{0}

\section{Further information about predictor variables}
\label{sec:SI_data}

As said in the main text, we have considered 16 socio-demographic variables, 7 mobility variables, and 8 environmental variables. All the variables are calculated at county-level. 
Before fitting the GLM model, all the predictive variables have been normalized to zero mean and unit standard deviation, while the output variable (number of deaths) is predicted in logarithmic form.
In the following subsections we provide further information about these variables, reporting results concerning the intraclass correlation for the three families of data, and graphical representations of the variables distributions.

\begin{figure}[b!]
    \centering
    \begin{subfigure}{.42\textwidth}
        \includegraphics[width=\textwidth]{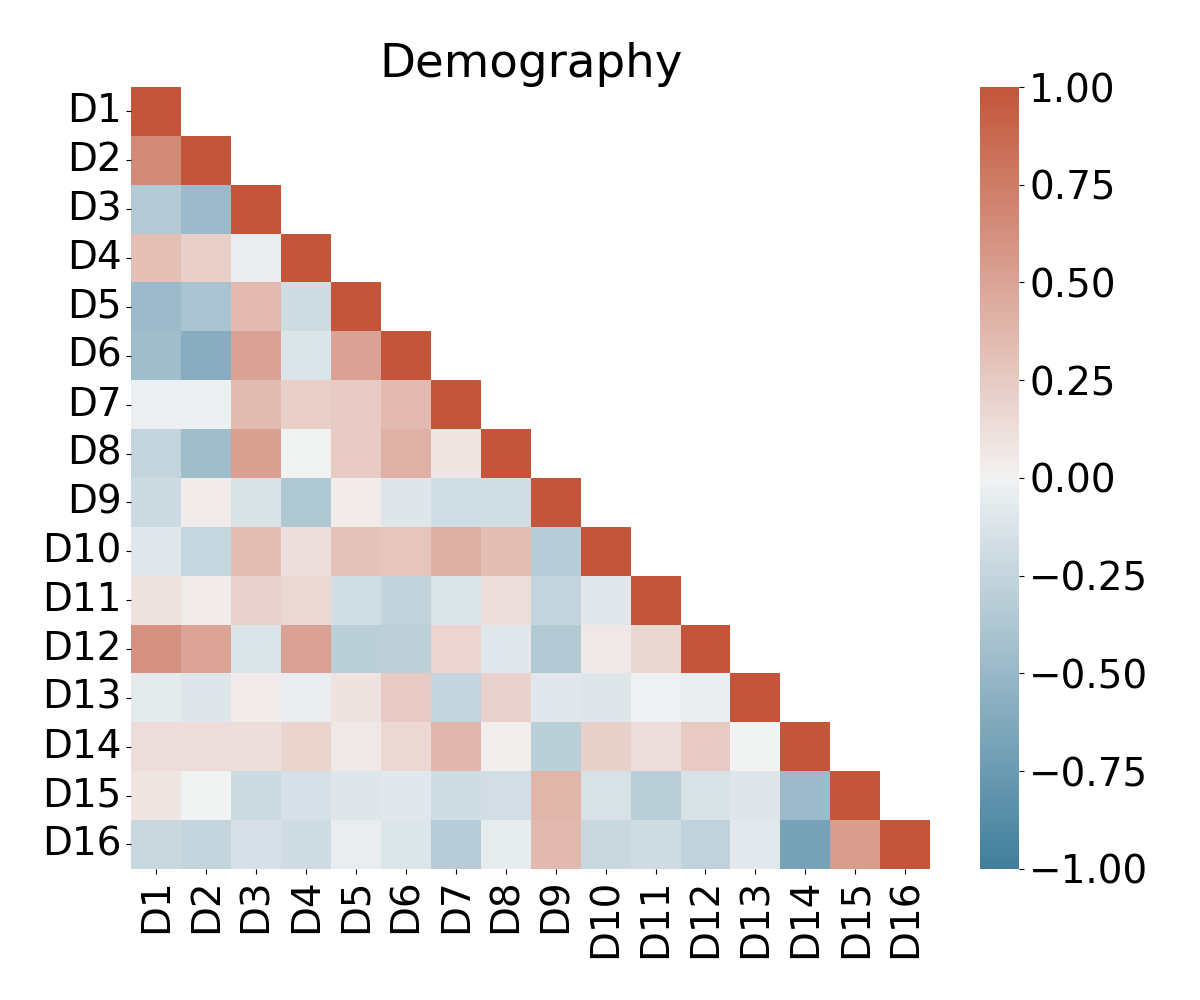}
        \caption{}\label{fig:feature_correlation_dem}
    \end{subfigure}
    \begin{subfigure}{.42\textwidth}
        \includegraphics[width=\textwidth]{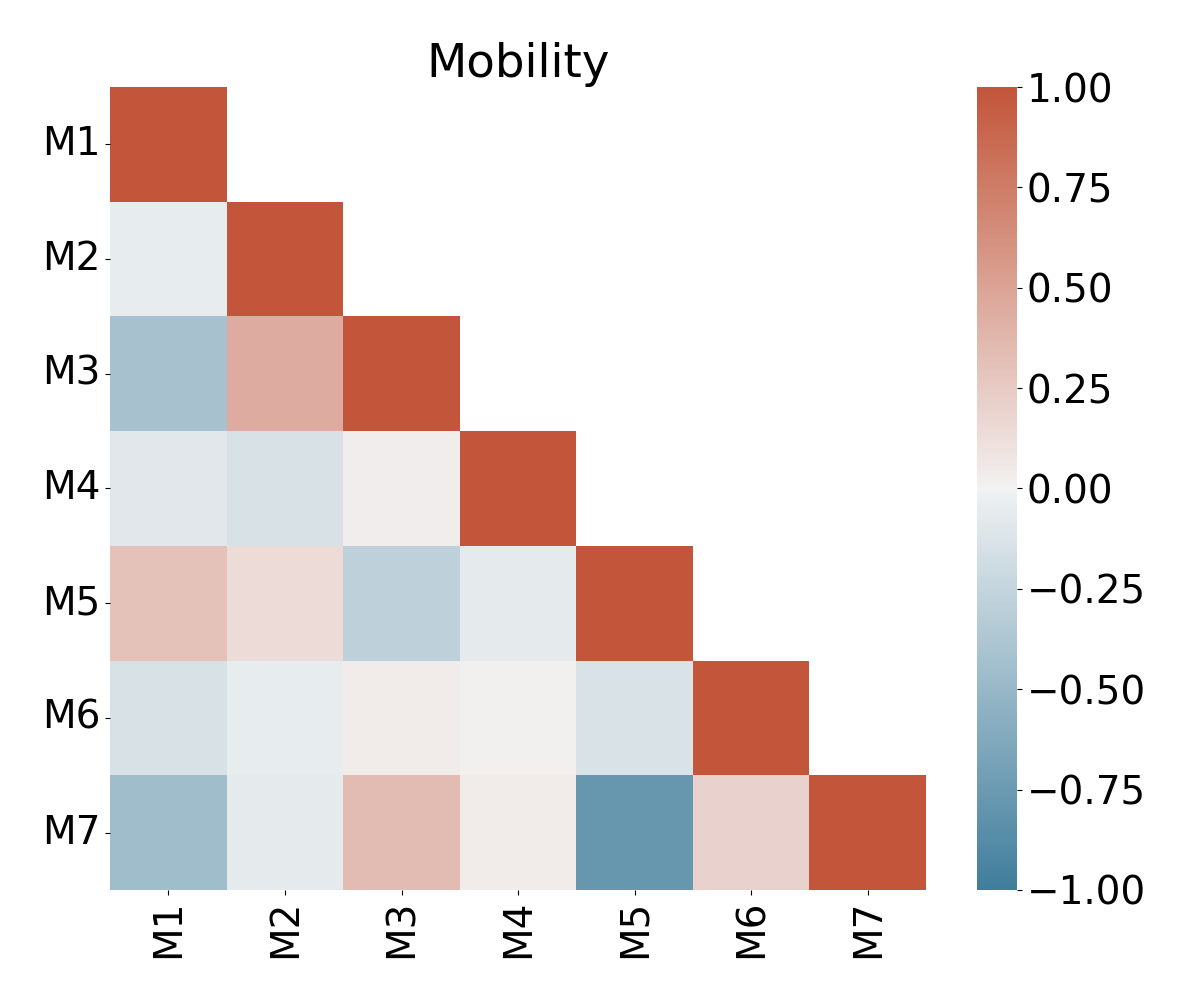}
        \caption{}\label{fig:feature_correlation_mob}
    \end{subfigure}
    \begin{subfigure}{.42\textwidth}
        \includegraphics[width=\textwidth]{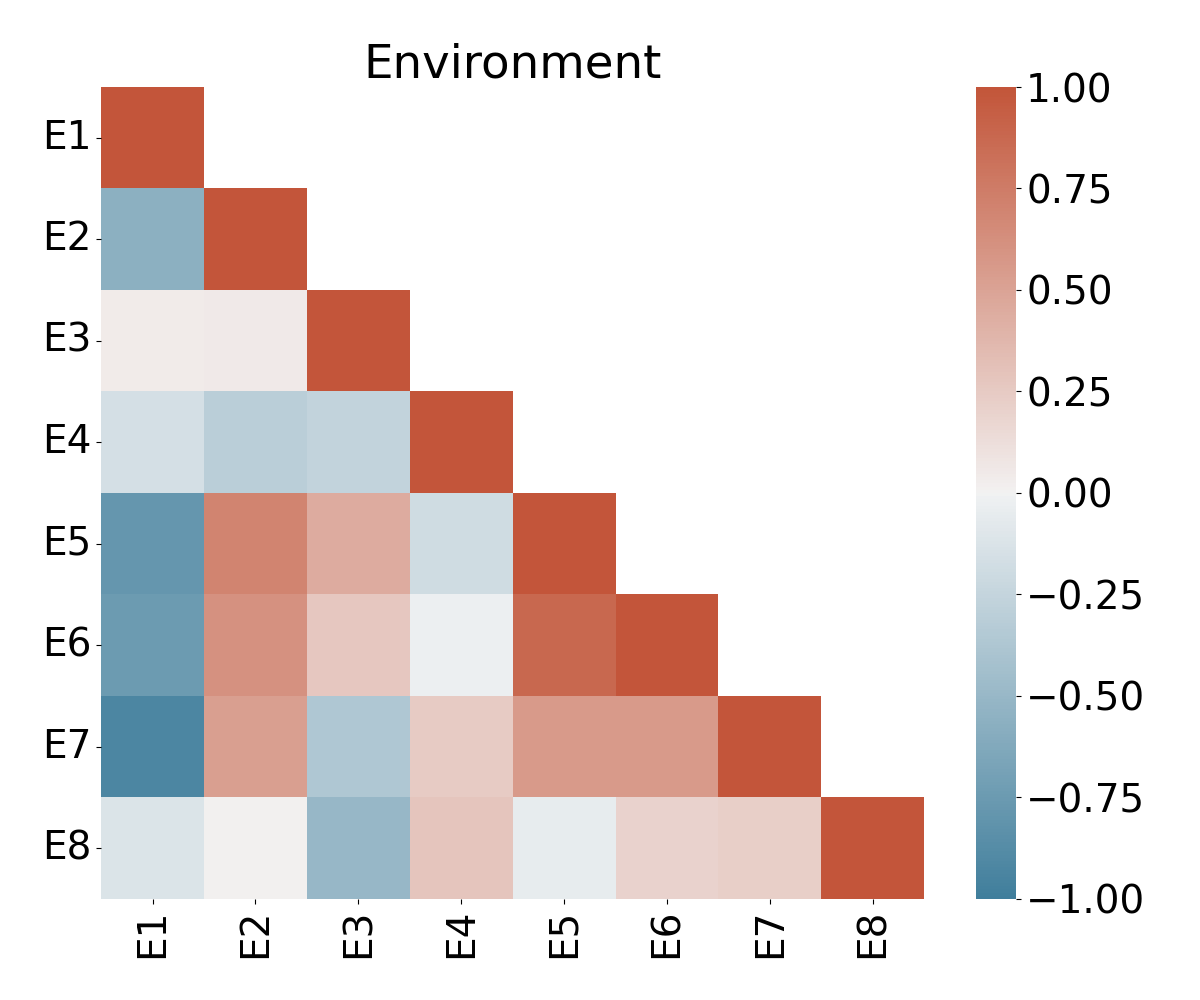}
        \caption{}\label{fig:feature_correlation_env}
    \end{subfigure}

    \caption{Correlation within each group of features: Socio-demographic (Fig.~\ref{fig:feature_correlation_dem}), Mobility (Fig.~\ref{fig:feature_correlation_mob}), and Environmental (Fig.~\ref{fig:feature_correlation_env}). Apart from some environmental variables, there are no significant correlations among the chosen predictors.} 
    \label{fig:feature_correlation}
\end{figure}

\subsection{Correlation analysis}
\label{sec:corr_analysis}
% Correlation analysis
To better understand the results from the generalized linear model, a Pearson product-moment correlation within each group of variables is depicted in Fig. \ref{fig:feature_correlation} (variables are reported with their own identification code introduced in Table \ref{tab:features}). As one can see, apart from some variables in the environmental group (e.g., mean temperature and O$_3$ are strongly negatively correlated, while specific humidity mean value and standard deviation are strongly positively correlated), there are not significant correlations among variables. Hence, we do not identify multi-collinearity issues in applying regression models using the entire proposed set of predictors. 

\subsection{Work time evolution during the first pandemic wave}
\label{sec:time_evolution_wt}

\begin{figure}[b!]
    \centering
    \includegraphics[width=.7\textwidth]{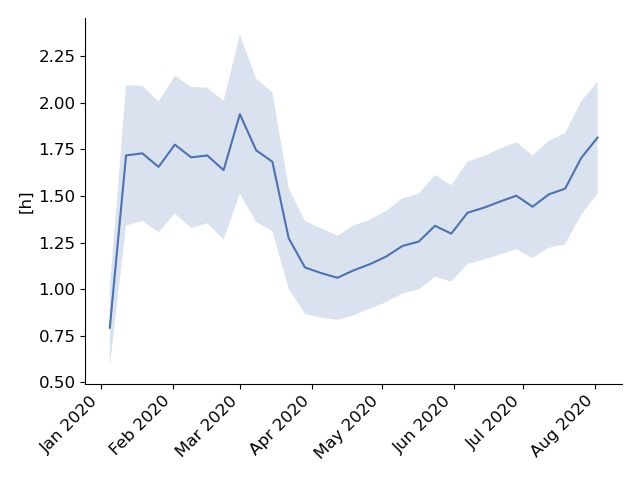}
\caption{Time evolution of the variable Work time ($M_2$). The plot shows the weekly aggregation of the mean over all the counties (solid line) and the corresponding standard deviation (shaded area).} 
    \label{fig:time_series_work_time}
\end{figure}

To improve the understanding of the effect of the time spent at work (variable Work time, $M_2$), we visualize in Figure~\ref{fig:time_series_work_time} its time evolution over the considered time. The data is obtained by first computing the mean and standard deviation, over all counties, of the daily quantity. The figure reports the weekly average of these mean (solid line) and standard deviation (shaded area).

A sharp drop of the time spent at work is visible during March-April 2020, followed by a steady but slow increase in the next months. The value of this variable in the final time (August 2020) is still below its pre-pandemic level.

\subsection{State-level data distribution}
\label{sec:state_level_dist}

\begin{figure}[!hp]
    \centering
    \includegraphics[width=1.0\textwidth]{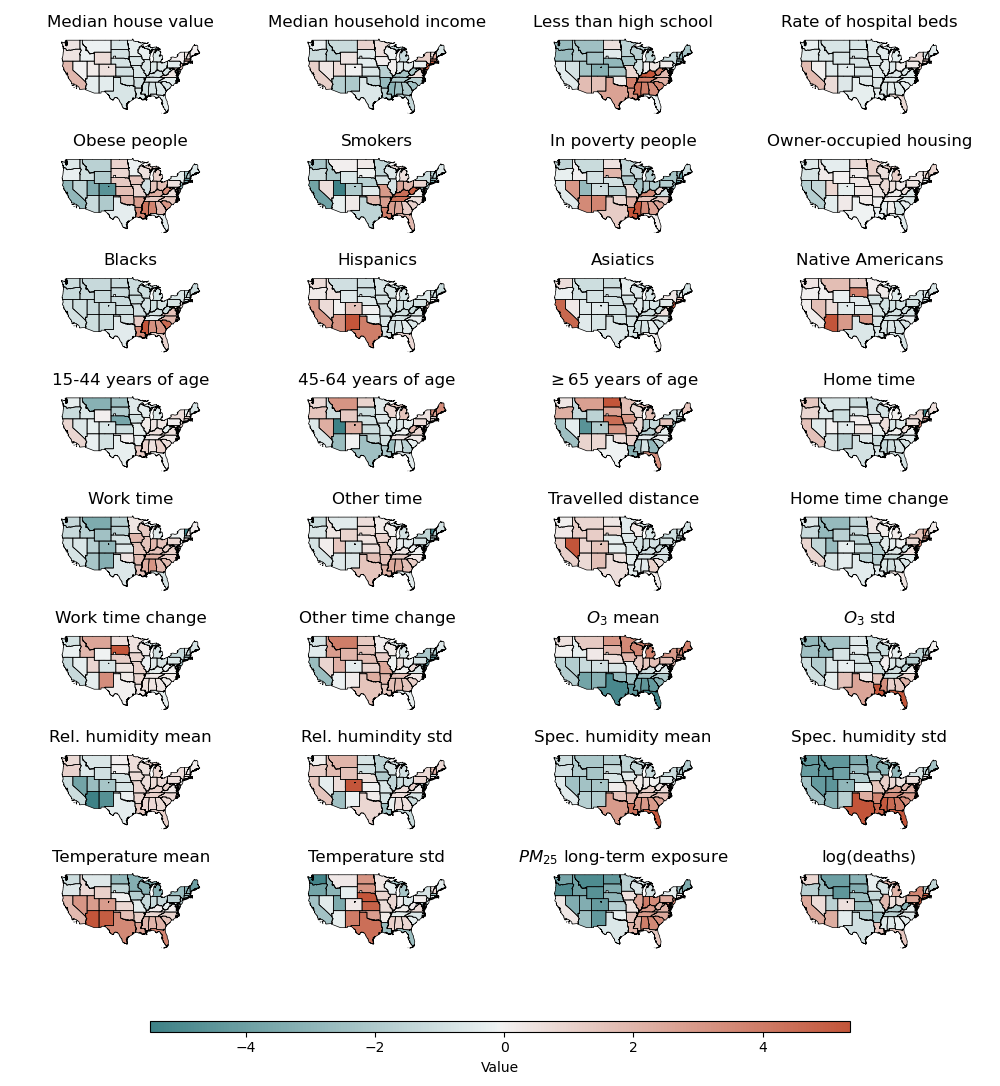}
        \caption{State-level distribution of the data. Each variable is normalized to have zero mean and unit standard deviation.}
    \label{fig:data_distr_per_state}
\end{figure}

The county-level distribution of COVID-19 deaths and of each predictor variable is reported in Fig. \ref{fig:data_distr_per_state}. Each variable is normalized to have zero mean and unit standard deviation.

\section{Fraction of Variance Unexplained for different combinations of variables groups}
\label{sec:table_map}

With reference to the maps reported in Fig. \ref{fig:maps_fvu_states}, we report in Table \ref{tab:FVU} the detailed results of the modelling analysis in terms of Fraction of Variance Unexplained (FVU). These numbers allow the reader to have a more detailed view of our results, thus better understanding the relevance of the demographic variables. Indeed, one can appreciate that when considering single groups, the demographic one is associated with the lower FVU values for all the US states. When considering combinations of groups, the groups containing demographic variables have lower FVU values than the mobility-environmental group.  

The FVU is the ratio between the Mean Squared Error (MSE) and the variance of the data. In details, given $n$ observed target values $y:=(y_1, \dots, y_n)$, and the corresponding predictions $\tilde y:=(\tilde y_1, \dots, \tilde y_n)$ provided by the GLM model, the FVU value is computed as
\begin{equation*}
    \mathrm{FVU} := \frac{\mathrm{MSE}(y,\tilde y)}{\mathrm{STD}(y)},
\end{equation*}
where 
\begin{equation*}
\mathrm{MSE}(y,\tilde y) := \frac{1}{n}\sum_{i=1}^n\left|y_i - \tilde y_i \right|^2, \quad\quad
\mathrm{STD}(y) := \frac{1}{n} \sum_{i=1}^n \left|y_i - \mathrm{MEAN}(y)\right|^2,
\quad\quad
\mathrm{MEAN}(y) = \frac{1}{n} \sum_{i=1}^n y_i.
\end{equation*}

\section{Analysis of the coefficients in the global generalized linear model}
\label{sec:si_beta_global_model}

\begin{figure}[htp!]
    \centering
    \includegraphics[height=.87\textheight]{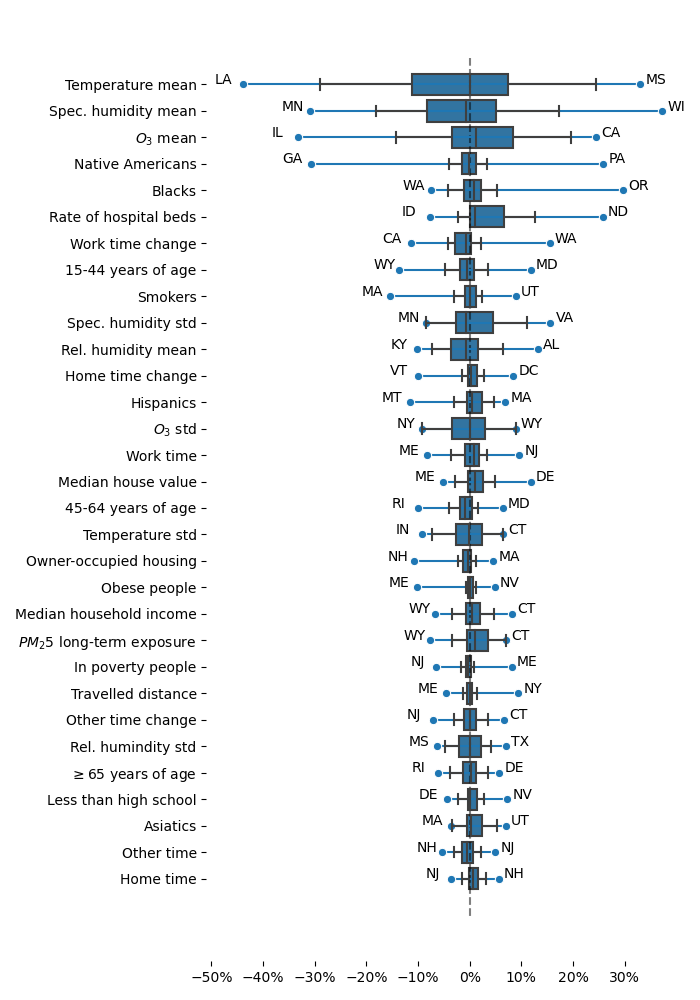}
\caption{Generalized Linear Model’s $\beta$ coefficients showing that Demographic, Mobility, and Environment features do not play the same role in all states. We highlight with blue dots the minimum and maximum
coefficient for each feature, and the related states.}
    \label{fig:beta_coeffs}
\end{figure}

In this section we report few additional details concerning the results of the GLM approach discussed in Sec. \ref{sec:results}. Here, we turn again our attention to the standardized $\beta$ coefficients that reveal how features correlate with COVID-19 mortality in each US state. The analysis is performed considering the complete model with all the variables in the three groups (the column \emph{All} in Table \ref{tab:FVU}). To properly compare the coefficients and consider the variability of number of deaths across the states, all the coefficients for every single state have been normalized such that they sum to one. Differently to what was shown and discussed in the main body of this research, here we focus on the minimum and maximum values of the $\beta$ coefficients attained by the US states. Hence, this analysis extends what was reported in Sec. \ref{sec:state_examples}.
The results are reported in Table \ref{tab:beta_coeffs} and depicted in Fig. \ref{fig:beta_coeffs} using box plots displaying interquartile ranges (the boxes reporting the spread difference between the 75th and 25th percentiles of the data) and outliers. In Fig. \ref{fig:beta_coeffs}, the variables have been sorted such that the ranges between minimum and maximum values are decreasing (which does not mean that also the interquartile ranges, i.e., the spread of the boxes, are decreasing monotonically in the same fashion). It is noteworthy that the $\beta$ coefficients greatly vary  across states, as one can appreciate by observing that all the reported variables negatively correlate with COVID-19 mortality in some states and positively correlate in other.
The reader can appreciate once again how the mobility variables play a relevant role in New Jersey, and this result was already discussed in Sec. \ref{sec:state_examples}.

\clearpage

\begin{table}[htp!]
\centering
\begin{adjustbox}{height=.445\textheight}
\begin{tabular}{lrrrrrrr}
\toprule
& \multicolumn{7}{c}{Fraction of Variance Unexplained [\%]}\\
State Code	&  Dem. 	&  Mob. 	&  Env. 		& Mob. + Env. & Dem + Env. & Dem. + Mob.  &  All 		   \\
\midrule                                                                                       
   AL &   	12.478 &   	78.106 &   	85.453 &   33.301	&    	9.405	&    	6.164	& 		5.570        \\
   AR &   	20.995 &   	67.769 &   	96.678 &   36.544	&    	12.152	&    	12.243	& 		8.663        \\
   AZ & 	$\sim 0$&   2.874 &   	13.441 &   $\sim 0$&    	$\sim 0$&    	$\sim 0$& 		$\sim 0$     \\
   CA &    	0.528 &   	10.761 &   	60.133 &   0.642	&    	0.035	&    	0.076	& 		0.018        \\
   CO &    	8.059 &   	38.706 &   	18.393 &   13.219	&    	2.268	&    	1.874	& 		0.437        \\
   CT & 	$\sim 0$& 	$\sim 0$& 	$\sim 0$&  $\sim 0$&    	$\sim 0$&    	$\sim 0$& 		$\sim 0$     \\
   DC & 	$\sim 0$& 	$\sim 0$& 	$\sim 0$&  $\sim 0$&    	$\sim 0$&    	$\sim 0$& 		$\sim 0$     \\
   DE & 	$\sim 0$& 	$\sim 0$& 	$\sim 0$&  $\sim 0$&    	$\sim 0$&    	$\sim 0$& 		$\sim 0$     \\
   FL &    	2.953 &   	15.524 &   	36.754 &   6.905	&    	1.271	&    	2.183	& 		0.947        \\
   GA &   	22.960 &   	33.147 &   	62.288 &   34.210	&    	21.199	&    	22.862	& 		19.104       \\
   IA &   	15.767 &   	38.838 &   	70.016 &   14.268	&    	6.811	&    	7.436	& 		3.359        \\
   ID &    	2.860 &   	16.267 &   	24.884 &   2.856	&    	25.252	&    	30.243	& 		$\sim 0$     \\
   IL &    	0.020 &    	1.558 &    	1.306 &    0.316	&    	0.015	&    	0.017	& 		0.012        \\
   IN &    	4.081 &   	88.887 &   	91.800 &   74.146	&    	2.996	&    	4.090	& 		2.087        \\
   KS &    	0.912 &    	8.469 &   	70.444 &   6.349	&    	0.719	&    	0.820	& 		0.610        \\
   KY &    	6.134 &   	88.162 &   	78.923 &   22.345	&    	5.163	&    	5.326	& 		3.474        \\
   LA &    	2.579 &   	20.932 &   	69.601 &   19.059	&    	2.092	&    	2.246	& 		1.835        \\
   MA & 	$\sim 0$&   2.425 &    	6.739 &    $\sim 0$&    	$\sim 0$&    	$\sim 0$& 		$\sim 0$     \\
   MD &    	0.242 &   	14.340 &   	22.555 &   0.802	&    	$\sim 0$&    	$\sim 0$& 		$\sim 0$     \\
   ME & 	$\sim 0$&   3.621 &    	6.958 &    $\sim 0$&    	$\sim 0$&    	$\sim 0$& 		$\sim 0$     \\
   MI &    	0.352 &   	10.444 &   	20.662 &   1.057	&    	0.073	&    	0.106	& 		0.044        \\
   MN &    	0.418 &   	11.411 &   	10.917 &   0.859	&    	0.136	&    	0.191	& 		0.104        \\
   MO &    	0.159 &    	5.774 &   	21.648 &   0.997	&    	0.098	&    	0.132	& 		0.078        \\
   MS &   	58.499 &   	89.067 &   	84.862 &   56.008	&    	39.513	&    	43.086	& 		16.281       \\
   MT &    	0.162 &   	88.898 &   	72.027 &   32.231	&    	40.161	&    	50.673	& 		$\sim 0$     \\
   NC &   	31.858 &   	50.274 &   	72.869 &   40.409	&    	28.177	&    	27.524	& 		24.800       \\
   ND &    	0.040 &    	2.574 &    	8.577 &    0.348	&    	$\sim 0$&    	$\sim 0$& 		14.610       \\
   NE &    	6.441 &   	47.897 &   	55.200 &   7.341	&    	2.191	&    	1.713	& 		1.574        \\
   NH & 	$\sim 0$&   0.003 &   	11.323 &   $\sim 0$&    	$\sim 0$&    	$\sim 0$& 		$\sim 0$     \\
   NJ &    	1.517 &   	23.709 &    3.828 &    1.156	&    	$\sim 0$&    	$\sim 0$& 		$\sim 0$     \\
   NM &    	0.263 &   	24.828 &   	13.261 &   0.030	&    	0.012	&    	0.001	& 		0.498        \\
   NV & 	$\sim 0$&   0.041 &    	0.016 &    0.092	&    	$\sim 0$&    	$\sim 0$& 		$\sim 0$     \\
   NY &    	0.148 &    	9.382 &    	5.485 &    2.610	&    	0.092	&    	0.091	& 		0.036        \\
   OH &   	15.173 &   	60.530 &   	63.874 &   38.785	&    	4.467	&    	8.995	& 		3.649        \\
   OK &    	7.123 &   	21.045 &   	73.955 &   8.862	&    	4.716	&    	2.986	& 		2.037        \\
   OR &    	0.312 &    	1.775 &    	7.109 &    0.126	&    	1.893	&    	5.630	& 		1.267        \\
   PA &    	4.241 &   	18.155 &    6.292 &    4.840	&    	1.165	&    	1.001	& 		0.472        \\
   RI & 	$\sim 0$& 	$\sim 0$& 	$\sim 0$&  $\sim 0$&    	$\sim 0$&    	$\sim 0$& 		$\sim 0$     \\
   SC &   	20.726 &   	78.339 &   	65.731 &   24.579	&    	5.263	&    	11.693	& 		4.084        \\
   SD &    	0.845 &    	2.035 &   	45.247 &   0.342	&		7.984	&		7.977	& 		$\sim 0$     \\
   TN &    	3.104 &   	50.406 &   	57.950 &   4.412	&    	2.350	&    	2.253	& 		1.348        \\
   TX &  	10.499 &   	33.472 &   	96.568 &   18.167	&    	5.991	&    	4.364	& 		3.320        \\
   UT &    	0.009 &    	2.048 &    	0.271 &    41.247	&    	0.054	&    	0.206	& 		$\sim 0$     \\
   VA &   	10.412 &   	65.557 &   	73.774 &   56.888	&    	5.240	&    	7.306	& 		3.559        \\
   VT & 	$\sim 0$&   0.056 &    	0.420 &    $\sim 0$&    	$\sim 0$&    	$\sim 0$& 		$\sim 0$     \\
   WA &    	0.285 &    	3.396 &   	53.627 &   1.607	&    	0.067	&	    0.137	& 		0.012        \\
   WI &    	0.568 &   	22.354 &   	39.003 &   3.994	&    	0.339	&    	0.330	& 		0.185        \\
   WV &    	7.028 &   	90.108 &   	86.925 &   67.066	&    	5.667	&    	5.248	& 		4.168        \\
   WY &   	40.403 & 	$\sim 0$& 	69.499 &   40.870	&    	$\sim 0$&    	$\sim 0$& 		$\sim 0$     \\
\bottomrule   
\end{tabular}
\end{adjustbox}
\caption{Fraction of Variance Unexplained for models considering one group of variables (columns from 2 to 4),  the combination of two groups of variables (columns from 5 to 7), and all the variables together (last column). Values $\sim 0$ are lower than $5\cdot10^{-5}\%$.}
\label{tab:FVU}
\end{table}

\newpage

\begin{table}[!htp]
\centering
\begin{tabular}{llclc}
\toprule
& \multicolumn{2}{c}{\emph{Minimum $\beta$ case}} & \multicolumn{2}{c}{\emph{Maximum $\beta$ case}}\\
               Variable & State & $\beta$ & State &  $\beta$ \\
\midrule
Median house value 				&        ME & -0.051 &     DE &  0.118 \\
Median household income 		&        WY & -0.068 &     CT &  0.083 \\
Less than high school 			&        DE & -0.043 &     NV &  0.073 \\
Rate of hospital beds 			&        ID & -0.077 &     ND &  0.257 \\
Obese people 					&        ME & -0.102 &     NV &  0.048 \\
Smokers 						&        MA & -0.154 &     UT &  0.090 \\
PM$_{2.5}$ long-term exposure 	&        WY & -0.078 &     CT &  0.070 \\
Below Poverty Line 				&        NJ & -0.065 &     ME &  0.082 \\
Owner-occupied housing 			&        NH & -0.108 &     MA &  0.046 \\
Black 							&        WA & -0.075 &     OR &  0.296 \\
Hispanic 						&        MT & -0.115 &     MA &  0.068 \\
Asiatic 						&        MA & -0.037 &     UT &  0.070 \\
Native Americans 				&        GA & -0.308 &     PA &  0.259 \\
15-44 years of age 				&        WY & -0.136 &     MD &  0.119 \\
45-64 years of age 				&        RI & -0.101 &     MD &  0.064 \\
$\ge$65 years of age 			&        RI & -0.062 &     DE &  0.057 \\
Home time 						&        NJ & -0.036 &     NH &  0.056 \\
Work time 						&        ME & -0.083 &     NJ &  0.096 \\
Other time 						&        NH & -0.053 &     NJ &  0.050 \\
Travelled distance 				&        ME & -0.046 &     NY &  0.095 \\
Home time change 				&        VT & -0.100 &     DC &  0.085 \\
Work time change 				&        CA & -0.114 &     WA &  0.155 \\
Other time change 				&        NJ & -0.072 &     CT &  0.066 \\
O$_3$ mean 						&        IL & -0.333 &     CA &  0.245 \\
O$_3$ std 						&        NY & -0.092 &     WY &  0.089 \\
Rel. Humidity mean 				&        KY & -0.103 &     AL &  0.132 \\
Rel. Humidity  std 				&        MS & -0.064 &     TX &  0.070 \\
Spec. Humidity mean 			&        MN & -0.310 &     WI &  0.373 \\
Spec. Humidity std 				&        MN & -0.085 &     VA &  0.156 \\
Temperature mean 				&        LA & -0.440 &     MS &  0.330 \\
Temperature std 				&        IN & -0.093 &     CT &  0.064 \\

\bottomrule
\end{tabular}
\caption{Maximum and minimum $\beta$ coefficients from the generalized linear model for each variable in the input dataset. Note that all the considered variables vary from negative to positive correlations across US states. Moreover, there is not a preeminent variable from one group, but variables from all three groups seem to be relevant in different states.}
\label{tab:beta_coeffs}
\end{table}

\end{document}